\documentclass[american,catalan,english]{article}
\usepackage[T1]{fontenc}
\usepackage[latin9]{inputenc}
\usepackage{amsmath}
\usepackage{amssymb}
\usepackage{epsfig,graphics}
\usepackage{hhline,longtable}

\makeatletter

\providecommand{\tabularnewline}{\\}

\newtheorem{theorem}{Theorem}

\newtheorem{corollary}[theorem]{Corollary}

\newtheorem{example}[theorem]{Example}

\newtheorem{proposition}[theorem]{Proposition}

\newenvironment{proof}[1][Proof]{\noindent\textbf{#1.} }{\ \rule{0.5em}{0.5em}}

\makeatother

\usepackage{babel}

\begin{document}

\title{Methods to distinguish between polynomial and exponential tails}

\author{Joan del Castillo \\
 Universitat Autonoma de Barcelona, Spain \and Jalila Daoudi
\\
 Universitat Autonoma de Barcelona, Spain \and Richard Lockhart
\\
Simon Fraser University, Canada}

\date{ }
\maketitle
\begin{abstract}
In this article two methods to distinguish between polynomial and
exponential tails are introduced. The methods are mainly based on
the properties of the residual coefficient of variation for the exponential
and non-exponential distributions. A graphical method, called CV-plot,
shows departures from exponentiality in the tails. It is, in fact,
the empirical coefficient of variation of the conditional excedance
over a threshold. The plot is applied to the daily log-returns of
exchange rates of US dollar and Japan yen.

New statistics are introduced for testing the exponentiality of tails
using multiple thresholds. Some simulation studies present the critical
points and compare them with the corresponding asymptotic critical
points. Moreover, the powers of new statistics have been compared
with the powers of some others statistics for different sample size. 
\end{abstract}


\section{Introduction}

Since Balkema-DeHaan (1974) and Pickands (1975), it has been well
known that the conditional distribution of any random variable over
a high threshold --- what is known in reliability as the residual
life --- has approximately a generalized Pareto distribution (GPD).
The exponential distribution is a particular case that appears between
compact support distributions and heavy-tailed distributions, in GPD.
Applications of extreme value theory to risk management in finance
and economics are now of increasing importance. The GPD has been used
by many authors to model excedances in several fields such as hydrology,
insurance, finance and environmental science, see McNeil et al. (2005),
Finkenstadt and Rootzén (2003), Coles (2001) and Embrechts et al.
(1997).

It is especially important for applications to distinguish between
polynomial and exponential tails. Often, the methodology is based
on graphical methods to determine the threshold where the tail begins,
see Embrechts et al. (1997) and Ghosh and Resnick (2010). In this
cases, multiple testing problem occurs when one considers a wide set
of thresholds.

The main objective of this paper is providing ways to distinguish
the behavior of tails, avoiding the multiple testing problems. The
methods are mainly based on the properties of the residual coefficient
of variation that is closely related to the likelihood functions of
the exponential and Pareto distributions, see Castillo and Puig (1999)
and Castillo and Daoudi (2009). The empirical coefficient of variation,
or equivalent statistics (e.g., Greenwood's statistic, Stephens $W_{s}$)
are omnibus tests used for testing exponentiality against arbitrary
increasing failure rate or decreasing failure rate alternatives. A
good description of these tests has been given by D'Agostino and Stephens
(1986).

A large number of tests for exponentiality have been proposed in the
literature. Montfort and Witter (1985) propose the maximum/median
statistic for testing exponentiality against GPD. Smith (1975) and
Gel, Miao and Gastwirth (2007) show that powerful tests of normality
against heavy-tailed alternatives \ are obtained using the average
absolute deviation from the median. Lee \emph{et al.} (1980) and Ascher
(1990) discuss tests based on the equation $E\left(X^{p}\right)/E\left(X\right)^{p}=\Gamma\left(1+p\right)$,
for some $p>0$, where $X$ is an exponential random variable. The
limit case, when $p$ tends to $0$, is studied in Mimoto and Zitikis
(2008), see also references therein. The case $p=2$ is equivalent
to the coefficient of variation test. Lee \emph{et al.} (1980) show
that in this case the power is poor testing against distributions
whose coefficient of variation is $1$ (the exponential case) as happens
testing against the absolute values of the Student distribution $t_{4}$.
Our methods based on a multivariate point of view are also useful
in this situation, since the exponential distribution is the unique
distribution with the residual coefficient of variation over any threshold
equal to $1$; see Sullo and Rutherford (1977), Gupta (1987) and Gupta
and Kirmani (2000).

In Section 2 the asymptotic distribution of the residual coefficient
of variation is studied as a random process in terms of the threshold.
This provides a clear graphical method, called a CV-plot, for assessing
departures from exponentiality in the tails. The qualitative behavior
of the CV-plot is made more precise in Section 3. The plot is applied
to the daily log-returns of exchange rates of US dollar and Japan
yen.

New statistics are introduced for testing the exponentiality of tails
using multiple thresholds in Section 4. Some simulation studies present
the critical points and compare them with the corresponding asymptotic
critical points.

In Section 5, the powers of new statistics have been compared with
the powers of some others statistics against heavy-tailed alternatives,
given by Pareto and absolute values of the Student distributions,
for different sample size.

\section{The residual coefficient of variation}

Let $X$ be a continuous non negative random variable with distribution
function $F\left(x\right)$. For any threshold, $t>0$, the distribution
function of threshold excedances, $\left(X-t\mid X>t\right)$, denoted
$F_{t}\left(x\right)$, is defined by \[
1-F_{t}\left(x\right)=\frac{1-F\left(x+t\right)}{1-F\left(t\right)}.\]
The coefficient of variation (CV) of the conditional excedance over
a threshold, $t$, (the residual CV) is \[
CV\left(t\right)=Var\left(X-t\mid X>t\right)^{1/2}/E\left(X-t\mid X>t\right)\]
where $E\left[\cdot\right]$ and $\ Var\left[\cdot\right]$ denote
the expected value and the variance. The $CV\left(t\right)$ is independent
of scale parameters. It will be useful find the distribution of the
empirical $CV$ process for all values of $t$.

It is well known that the mean residual lifetime determines the distribution
for random variables. Gupta and Kirmani (2000) showed that mean residual
life is a function of the residual coefficient of variation, hence
it also characterize the distribution. In this context, generalized
Pareto distributions appear as the simple case in which the residual
coefficient of variation is a constant. Hence, from Pickands (1975)
and Balkema-DeHaan (1974), it is almost constant for a sufficiently
high threshold.

Denote $X1_{\left(X>t\right)}$ the random variable $X$ if it is
larger that $t$ and zero otherwise. Denote $\mu_{0}\left(t\right)=\Pr\left\{ X>t\right\} $
and $\mu_{k}\left(t\right)=E\left[X^{k}1_{\left(X>t\right)}\right]$,
$k>0$. Throughout this paper $\mu_{0}\left(t\right)>0,$ foll all
$t$, is assumed. Note that\[
\mu_{k}\left(t\right)=\mu_{0}\left(t\right)E\left(X^{k}\mid X>t\right).\]

Given a sample $\left\{ X_{j}\right\} $ of size $n$, let $n\left(t\right)=\sum_{j=1}^{n}1_{\left(X_{j}>t\right)}$
the number of excedance over a threshold, $t$. By the law of large
numbers, $n\left(t\right)/n$ converges to $\mu_{0}\left(t\right)$.
\emph{The empirical CV of the conditional excedance} is given by\begin{multline}
cv_{n}\left(t\right)=\frac{n\left(t\right)}{\sum_{j=1}^{n}\left(X_{j}-t\right)1_{\left(X_{j}>t\right)}}\times\\
\left[\frac{\sum_{j=1}^{n}X_{j}^{2}1_{\left(X_{j}>t\right)}}{n\left(t\right)}-\left(\frac{\sum_{j=1}^{n}X_{j}1_{\left(X_{j}>t\right)}}{n\left(t\right)}\right)^{2}\right]^{1/2}\label{cv1}\end{multline}

The $cv_{n}\left(t\right)$ is also independent of scale parameters,
since the mean and standard deviation have the same units.

\begin{proposition} The $cv_{n}\left(t\right)$ is a consistent estimator
of $CV\left(t\right)$, assuming finite second moment, since the limit
in probability of $cv_{n}\left(t\right)$, as $n$ goes to infinity
is\[
m_{cv}\left(t\right)=\frac{\sqrt{\mu_{2}\left(t\right)\mu_{0}\left(t\right)-\mu_{1}\left(t\right)^{2}}}{\mu_{1}\left(t\right)-t\mu_{0}\left(t\right)}=CV\left(t\right)\]

\begin{proof} Fixed $t$, as $n$ goes to infinity\[
\frac{1}{n\left(t\right)}\sum_{j=1}^{n}X_{j}^{k}1_{\left(X_{j}>t\right)}=\frac{n}{n\left(t\right)}\frac{1}{n}\sum_{j=1}^{n}X_{j}^{k}1_{\left(X_{j}>t\right)}\rightarrow\mu_{k}\left(t\right)/\mu_{0}\left(t\right)=E\left[X^{k}\mid X>t\right],\]
by the law of large numbers. Hence, the limit in probability of $cv_{n}\left(t\right)$\ is\[
\frac{\sqrt{\mu_{2}\left(t\right)/\mu_{0}\left(t\right)-\left(\mu_{1}\left(t\right)/\mu_{0}\left(t\right)\right)^{2}}}{\mu_{1}\left(t\right)/\mu_{0}\left(t\right)-t}=\frac{\sqrt{Var\left(X-t\mid X>t\right)}}{E\left(X-t\mid X>t\right)}\]
 \end{proof} \end{proposition}

Let us define the standardized $k$-th sampling moment of the conditional
excedance by\[
W_{k,n}\left(t\right)=\frac{1}{\sqrt{n}}\sum_{j=1}^{n}\left\{ X_{j}^{k}1_{\left(X_{j}>t\right)}-\mu_{k}\left(t\right)\right\} ,\]
hence,\begin{equation}
\sum_{j=1}^{n}X_{j}^{k}1_{\left(X_{j}>t\right)}=\sqrt{n}W_{k,n}\left(t\right)+n\mu_{k}\left(t\right).\label{as}\end{equation}
Note that normalizing constant $1/\sqrt{n}$ is used in order to have
$W_{k,n}\left(t\right)=O_{p}\left(1\right)$, with orders of convergence
in probability notation. The covariance of this random process is
given by \begin{align}
cov\left(W_{i,n}\left(s\right),W_{j,n}\left(t\right)\right) & =cov\left(X^{i}1_{\left(X>s\right)},X^{j}1_{\left(X>t\right)}\right)\notag\\
 & =\mu_{i+j}\left(s\vee t\right)-\mu_{i}\left(s\right)\mu_{j}\left(t\right),\label{cov}\end{align}

Throughout this paper the quantities $cv$ and $W_{k}$ among others
depend on $n$; \ wherever possible the dependence of quantities
on $n$ is suppressed for simplicity. Even the dependence on $t$
is dropped for $W_{k}=W_{k}\left(t\right)$ and $\mu_{k}=\mu_{k}\left(t\right)$,
in many places.

\begin{theorem} \label{p2}Let $X$ be a continuous non negative
random variable with finite fourth moment. Then, the following expansion
holds\begin{eqnarray}
\sqrt{n}\left(cv\left(t\right)-m_{cv}\left(t\right)\right)\! &\!\! =\!\! &\! \frac{\mu_{0}W_{2}}{2\left(\mu_{1}-t\mu_{0}\right)\!\sqrt{\mu_{2}\mu_{0}-\mu_{1}^{2}}}+\frac{\mu_{0}\left(t\mu_{1}-\mu_{2}\right)W_{1}}{\left(\mu_{1}-t\mu_{0}\right)^{2}\!\sqrt{\mu_{2}\mu_{0}-\mu_{1}^{2}}}+\!\notag\label{cv2}\\
\\ \!&\!  & +\frac{\left(-2t\mu_{1}^{2}+t\mu_{0}\mu_{2}+\mu_{1}\mu_{2}\right)W_{0}}{2\left(\mu_{1}-t\mu_{0}\right)^{2}\sqrt{\mu_{2}\mu_{0}-\mu_{1}^{2}}}+O_{p}\left(\frac{1}{\sqrt{n}}\right).\notag\end{eqnarray}
 \end{theorem}

\begin{proof} The expression $\left(\ref{cv1}\right)$ in terms of
$W_{k}=W_{k,n}\left(t\right)$ is
\begin{multline}
cv\left(t\right)=\frac{\mu_{0}\left(t\right)+W_{0}/\sqrt{n}}{\mu_{1}\left(t\right)+W_{1}/\sqrt{n}-t\left(\mu_{0}\left(t\right)+W_{0}/\sqrt{n}\right)}\times\\
\left[\frac{\mu_{2}\left(t\right)+W_{2}/\sqrt{n}}{\mu_{0}\left(t\right)+W_{0}/\sqrt{n}}-\left(\frac{\mu_{1}\left(t\right)+W_{1}/\sqrt{n}}{\mu_{0}\left(t\right)+W_{0}/\sqrt{n}}\right)^{2}\right]^{1/2}\label{cv3}\end{multline}
Let $w_{k}=W_{k}/\sqrt{n}=O_{p}\left(1/\sqrt{n}\right)$, since $W_{k}=O_{p}\left(1\right).$
Then, let us replace $w_{k}$ in $\left(\ref{cv3}\right)$. Taking
a Taylor expansion of $\sqrt{n}\left(cv\left(t\right)-m_{cv}\left(t\right)\right)$
with respect to $w_{k}$ near zero the result follows. \end{proof}

\begin{example} \label{pe}Let $X$ be a random variable with an
exponential distribution with mean $\mu$. Conditional moments of
$X$, $\mu_{k}\left(t\right)$, can be obtained from the conditional
moments of the exponential distribution of mean $1,\mu_{k}^{1}\left(t\right)$
by\[
\mu_{k}\left(t\right)=\mu^{k}\mu_{k}^{1}\left(t/\mu\right)\]
where \begin{eqnarray*}
\mu_{0}^{1}\left(t\right) & \!\!\!\!=\!\!\!\! & e^{-t},\qquad\mu_{1}^{1}\left(t\right)=e^{-t}\left(1+t\right),\qquad\mu_{2}^{1}\left(t\right)=e^{-t}\left(2+t\left(2+t\right)\right)\\
\mu_{3}^{1}\left(t\right) &\!\!\!\!=\!\!\!\! & e^{-t}\left(6+t\left(6+t\left(3+t\right)\right)\right),\,\mu_{4}^{1}\left(t\right)=e^{-t}\left(24+t\left(24+t\left(12+t\left(4+t\right)\right)\right)\right).\end{eqnarray*}
In particular \[
m_{cv}\left(t\right)=1.\]
 \end{example}

In this Section several results on the convergence of random processes
are shown, in the sense of convergence of finite-dimensional distributions.
These results are sufficient for the applications given in Section
4.

If tightness is proved then weak convergence in the Skorokhod space
follows, but this will not be considered here.

\begin{corollary} \label{c1}Let $X$ be a random variable with exponential
distribution of mean $\mu$; then $\sqrt{n}\left(cv\left(t\right)-1\right)$
converges to a Gaussian process with zero mean and covariance function
given by\[
\rho\left(s,t\right)=\exp\left(\frac{s\wedge t}{\mu}\right)\text{.}\]
In particular\begin{equation}
\sqrt{n}\left(cv\left(0\right)-1\right)\overset{d}{\rightarrow}N\left(0,1\right)\text{,}\label{Green}\end{equation}
that corresponds to the asymptotic distribution of Greenwood's statistic.
\end{corollary}

\begin{proof} From Theorem $\ref{p2}$ and Example $\ref{pe}$ it
follows that\[
\sqrt{n}\left(cv\left(t\right)-1\right)=\left(W_{0},W_{1},W_{2}\right)\ a\left(t\right)+O_{p}\left(n^{-1/2}\right)\]
where\[
a\left(t\right)^{\prime}=(e^{t/\mu}\,\left(t^{2}+4\, t\,\mu+2\,{\mu}^{2}\right)/\left(2\,{\mu}^{2}\right),-e^{t/\mu}\,\left(t+2\,\mu\right)/{\mu}^{2},e^{t/\mu}/\left(2\,{\mu}^{2}\right))\]
Then, the covariance matrix of $W=\left(W_{0},W_{1},W_{2}\right)^{\prime}$,
from $\left(\ref{cov}\right)$ and Example \ref{pe}, assuming $s\leq t$,
is \[
\mathrm{cov}\left(W(s),W(t)\right)\equiv M\left(s,t\right)=\left(\mu_{i+j}\left(t\right)-\mu_{i}\left(s\right)\mu_{j}\left(t\right)\right)_{i,j=0,1,2}.\]
Some algebra shows\[
a\left(s\right)^{\prime}M\left(s,t\right)\ a\left(t\right)=\exp\left(s/\mu\right).\]
 \end{proof}

\begin{proposition} Let $X$ be a random variable with exponential
distribution of mean $\mu$; then using a new time scale, $\tau=\mu\log t$,
for $t\geq1$, the random process of $\sqrt{n}\left(cv\left(\tau\right)-1\right)$
converges to standard Brownian Motion. \end{proposition}

\begin{proof} From $\left(\ref{c1}\right)$, given $s,t\geq1$, \[
\rho\left(\mu\log s,\mu\log t\right)=\exp\left(\log s\wedge\log t\right)=s\wedge t\]
 \end{proof}

Corollary $\ref{c1}$ uses the same $n$ in $\sqrt{n}\left(cv\left(t\right)-1\right)$
for all $t$. The next result uses the sample size adapted to the
corresponding $t$.

\begin{corollary} Let $X$ be a random variable with an exponential
distribution, then $\sqrt{n\left(t\right)}\left(cv\left(t\right)-1\right)$
converges to a Gaussian process with zero mean and covariance function
given by \[
\exp\left(-\left\vert s-t\right\vert /\left(2\mu\right)\right).\]
This is the covariance function of the Ornstein-Uhlenbeck process,
the continuous time version of an $AR(1)$ process. It is a stationary
Markov Gaussian process. In particular, for any fixed $t$\begin{equation}
\sqrt{n\left(t\right)}\left(cv\left(t\right)-1\right)\overset{d}{\rightarrow}N\left(0,1\right)\text{.}\label{ex}\end{equation}
 \end{corollary}

\begin{proof} We remember that $n\left(t\right)/n$ converges to
$\mu_{0}\left(t\right)=\Pr\left\{ X>t\right\} >0$. Hence, if $n$
tends to infinity $n(t)$ tends to infinity too$.$We can write \[
\sqrt{n\left(t\right)}\left(cv\left(t\right)-1\right)=\sqrt{n\left(t\right)/n}\sqrt{n}\left(cv\left(t\right)-1\right).\]
From $\left(\ref{as}\right)$ and Example $\ref{pe}$, we have\[
\frac{n\left(t\right)}{n}=\exp\left(-t/\mu\right)+\frac{W_{0}}{\sqrt{n}}.\]
Then $\sqrt{n\left(t\right)}\approx\sqrt{n}\exp\left(-t/2\mu\right)$
and we have that \[
\exp\left(-s/2\mu\right)\exp\left(\frac{s\wedge t}{\mu}\right)\exp\left(-t/2\mu\right)=\exp\left(-\left\vert s-t\right\vert /\left(2\mu\right)\right).\]
 \end{proof}

\section{CV-plot}

Given a sample $\left\{ x_{k}\right\} $ of positive numbers of size
$n$, we denote by $\left\{ x_{\left(k\right)}\right\} $ the ordered
sample, so that $x_{\left(1\right)}\leq x_{\left(2\right)}\leq...\leq x_{\left(n\right)}$.
We denote by \emph{CV-plot} the representation of the empirical CV
of the conditional excedance $\left(\ref{cv1}\right)$, given by\begin{equation}
k\rightarrow cv\left(x_{\left(k\right)}\right)\text{.}\label{plot1}\end{equation}

The CV-plot does not depend on scale parameters, since the $cv_{n}\left(t\right)$
does not. That is, the CV-plots for samples $\left\{ x_{k}\right\} $
and $\left\{ \lambda\ x_{k}\right\} $ are the same, for any $\lambda>0$.
\ In order to have a reference for the behavior of $\left(\ref{plot1}\right)$,
pointwise error limits for these plots can be obtained for large samples
using $\left(\ref{ex}\right)$, from the null hypothesis of exponentiality.
In Section 4, pointwise error limits of the CV-plot are computed by
simulation for samples of several sizes. Then, the points are joined
by linear interpolation and plotted in the CV-plots.

Under regularity conditions, the conditional distribution of any random
variable over a high threshold is approximately GPD and this model
is characterized as the family of distributions with constant residual
CV, as has been said. Hence, the CV-plot can be a complement tool
to the Hill-plot or the ME-plot, which are used as diagnostics in
the extreme values theory, see Ghosh and Resnick (2010).

In order to illustrate the usefulness of the residual coefficient
of variation, we are going to examine the behavior of exchange rates
between the US dollar and the Japanese yen (JPY), from January 1,
1979 to December 31, 2003. The data set is available from OANDA Corporation
at http://www.oanda.com/ convert/fxhistory.

The \emph{daily returns} for the dollar price, $P_{k}$, are given
by\[
x_{k}=\log\left(P_{k}\right)-\log\left(P_{k-1}\right)\]
The daily returns are assumed to be independent here, as in the most
basic financial models. However, the theory may be extended even for
short-range correlations, see Coles (2001, chap. 5)

The set of positive returns is called the \emph{positive part }of
returns\emph{\ }and\emph{\ } the set of minus the negative returns
is called the \emph{negative part}. Both cases are samples of positive
random variables. From the 25 years considered we have 9131 daily
returns, 3840 of which are positive, 3642 negative and 1649 are equal
to zero.

In Figure 1, the plots (a) and (b) are the CV-plots of the $n=2000$
largest values for the positive and negative part of dollar/yen returns,
respectively. Pointwise 90\% limits around the line $cv=1$ are included,
the lowest sample size we consider is $20$, since not relevant information
comes from smaller samples. Since the basic model for returns is the
normal distribution, we will assume that the distribution has support
in $(0,\infty)$. \ Then, their threshold excedances, for large thresholds,
are very nearly Pareto distributed with parameter $\xi>0$ (Pickands,
1975). Some remarks arise from Figure 1. The plot (a) shows that the
process $\left(\ref{plot1}\right)$ for the positive part of dollar/yen
returns is always inside the pointwise limits for the exponential
distribution. Moreover, since we are only interested to test against
Pareto alternatives, we have to consider only upper bounds; thus the
pointwise level is $95\%$. Hence, the hypothesis that $CV=1$ can
be accepted and we can say that the tails decrease at an exponential
rate. Note that use of simultaneous confidence limits would make the
bounds wider, reinforcing our conclusion.

The plot (b) shows that the process $\left(\ref{plot1}\right)$ for
the negative part of dollar/yen returns is clearly outside the error
limits for the exponential distribution in most of the range. It seems
clear that we have to reject the hypothesis of exponentiality. However,
the coefficient of variations looks like a constant, approximately.
Hence, a Pareto distribution might be accepted for the sample.
\begin{figure}[h]
\begin{center}
\includegraphics{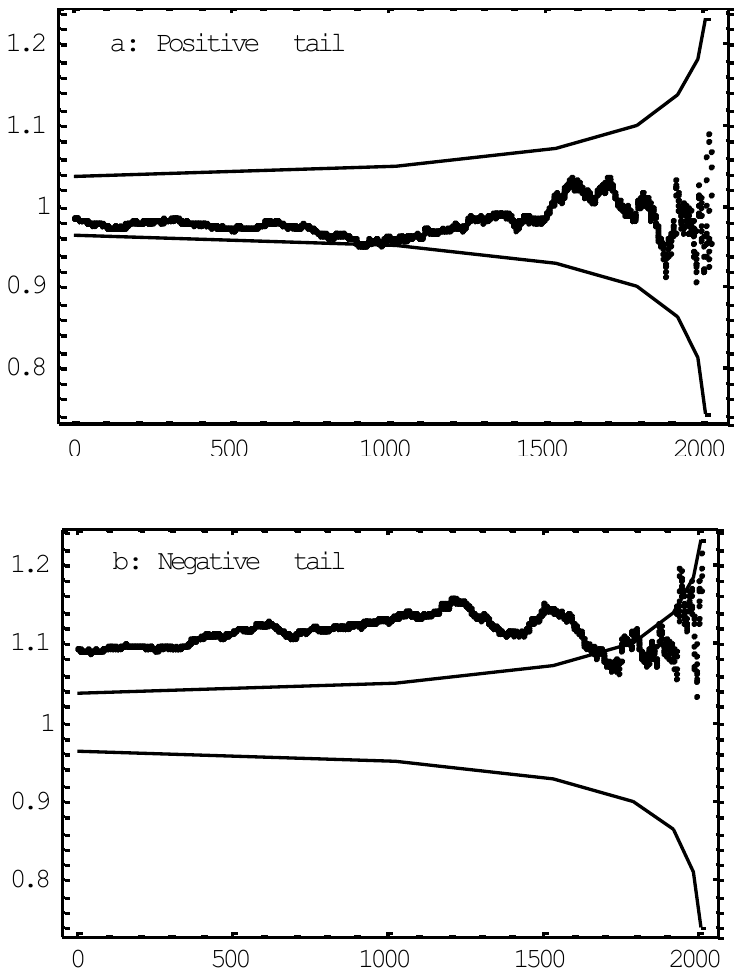}
\caption{
The plots (a) and (b) are the CV-plots of the $n = 2000$ largest values for the positive and negative parts of dollar/yen returns, respectively, with
pointwise 90{\%} error limits under the exponential distribution hypothesis.}
\end{center}
\end{figure}

\section{Testing exponentiality allowing multiple thresholds}

The CV-plot, explained in the last subsection, provides a clear graphical
method for assessing departures from exponentiality in the tails.
This qualitative behavior shall be made more precise here by introducing
new tests of exponential tails adapted to the present situation. The
tests are more powerful than most tests against the absolute values
of the Student distribution, as we will see in Section~\ref{powerestimates},
including the empirical coefficient of variation, or equivalent statistics
as Greenwood's statistic or Stephens $W_{s}$ (D'Agostino and Stephens,
1986). Our approach is the following:

Given a sample $\left\{ x_{j}\right\} $ from an exponential distribution,
for any set of thresholds $t_{0}<t_{1}<...<t_{m}$, let $n\left(t_{k}\right)$
be the number of events in $\left\{ x_{j}:x_{j}>t_{k}\right\} $,
and $cv\left(t_{k}\right)$ the empirical CV given by $\left(\ref{cv1}\right)$,
where $0\leq k\leq m$.

From $\left(\ref{ex}\right)$, asymptotically $n\left(t_{k}\right)\left(cv\left(t_{k}\right)-1\right)^{2}$
is distributed as a $\chi_{1}^{2}$ distribution. Let us consider
the statistic\begin{equation}
T=\sum_{k=0}^{m}n\left(t_{k}\right)\left(cv\left(t_{k}\right)-1\right)^{2}.\label{cd0}\end{equation}

Clearly the asymptotic expectation of $T$\ is $m+1$; however, its
asymptotic distribution is not $\chi_{m+1}^{2}$, since the random
variables $cv\left(t_{k}\right)$ are not independent. Its distribution
does not depend on scale parameters and it is straightforward to simulate
the distribution of $T$. It is important to note that lower values
for $T$ are expected under the null hypothesis of exponentiality,
when the expected values for $cv\left(t_{k}\right)$ are $1$. Hence,
high values for $T$ show departure from exponential tails.

The thresholds $\left\{ t_{k}\right\} $ can be arbitrary but some
practical simplicity is obtained by taking thresholds approximately
equally spaced, under the null hypothesis of exponentiality. The next
result shows a way of doing this.

\begin{proposition} \label{p4}If $X$ is a random variable with
exponential distribution of mean $\mu$, then \[
\Pr\left\{ X>\left(\mu\log2\right)k\right\} =1/2^{k}\]
 \end{proposition}

Given a sample $\left\{ x_{j}\right\} $ of size $n$ with exponential
distribution, the subsample of the last $n/2^{k}$ elements (assuming
that $n/2^{k}$ is integer) corresponds to the elements greater than
the order statistic $x_{\left(n-n/2^{k}\right)}$ and $x_{\left(0\right)}=0$,
$x_{\left(n/2\right)}$, $x_{\left(3\ n/4\right)}$, $x_{\left(7\ n/8\right)}$,
... are approximately equally spaced, from Proposition $\ref{p4}$.
\ 

For a general sample, the quantiles $q_{k}$ corresponding to the
last $n/2^{k}$ elements are considered $(q_{1}$ is the median, $q_{2}$
is the third quartile, ...$)$. From $\left(\ref{p4}\right)$, $q_{k}\approx\left(\mu\log2\right)k\approx x_{\left(n-n/2^{k}\right)}$.
Taking the set of thresholds corresponding to these sampling quantiles,
$\left(\ref{cd0}\right)$ became \begin{equation}
T_{m}=n\sum_{k=0}^{m}2^{-k}\left(cv\left(q_{k}\right)-1\right)^{2}\label{cd}\end{equation}

\subsection{Asymptotic distribution}

It is possible to write $\left(\ref{cd}\right)$ in the form $T_{m}$
$=V^{\prime}V$, where \[
V^{\prime}=\sqrt{n}\left[cv\left(q_{0}\right)-1,2^{-1/2}\left(cv\left(q_{1}\right)-1\right),...,2^{-m/2}(cv\left(q_{m}\right)-1)\right]\]

The asymptotic distribution of $T_{m}$ can be found from Corollary
$\ref{c1}$ in the following way. From Proposition $\ref{p4},$ we
have that $q_{k}\approx\left(\mu\log2\right)k$. Then, asymptotically,
the covariance matrix for $V$ is\[
\Sigma_{m}=\left(2^{-i/2}\rho\left(q_{i},q_{j}\right)2^{-j/2}\right)_{i,j=0,...,m}=\left(2^{-\left\vert i-j\right\vert /2}\right)_{i,j=0,...,m}\]

\begin{theorem} \label{p3}The asymptotic distribution of $T_{m}$
is $\sum_{0}^{m}\lambda_{i}Z_{i}^{2}$ with $Z_{i}$ distributed as
independent $N(0,1)$ and $\lambda_{i}$ the eigenvalues of $\Sigma_{m}$.
\end{theorem}

\begin{proof} From the central limit theorem $V$ is asymptotically
multivariate normal $N\left(0,\Sigma_{m}\right)$. Then, in a classical
argument, $\Sigma_{m}=A\ \Lambda\ A^{\prime}$ with $A$ an orthogonal
matrix and $\Lambda$ the diagonal matrix of the eigenvalues. It follows
that $V=A\ \Lambda^{1/2}Z$ with $Z$ asymptotically multivariate
normal with the identity as covariance matrix, $N\left(0,I\right)$.
Then $T_{m}$ $=V^{\prime}V=Z^{\prime}\Lambda\ Z=\sum_{0}^{m}\lambda_{i}Z_{i}^{2}$,
because $A$ is an orthogonal matrix. \end{proof}

\begin{example} For instance, for $m=2$,\[
\Sigma_{2}=\left(\begin{array}{ccc}
1 & 1/\sqrt{2} & 1/2\\
1/\sqrt{2} & 1 & 1/\sqrt{2}\\
1/2 & 1/\sqrt{2} & 1\end{array}\right)\]
and the eigenvalues are given by\[
\lambda_{0}=\left(5+\sqrt{17}\right)/4,\lambda_{1}=1/2,\lambda_{2}=\left(5-\sqrt{17}\right)/4\]
 \end{example}

Note also that for $m=0$, the asymptotic distribution of $T_{0}$
is simply a $\chi_{1}^{2}$ distribution. Numerical values of the
eigenvalues $\lambda_{i}$ are given in Table 1 for other small values
of $m$.

\subsection{Approximate critical points}

Simulation methods are now easily available to compute critical values
and p-values of $T_{m}$. However, the asymptotic distribution of
$T_{m}$, given by Theorem $\ref{p3}$, provides a way to compute
such p-values for large sample sizes without heavy simulation. For
instance, if the sample size is $n=2000$ and $m=3$, the direct method
needs samples of $2000$ exponential random numbers and the asymptotic
distribution only needs samples of $4$ normal random numbers.

Moreover, the asymptotic distribution of $T_{m}$, given by Theorem
$\ref{p3}$, can be approximated by $a+b\chi_{\nu}^{2}$, where $\chi_{\nu}^{2}$
has gamma distribution with parameters $\left(\nu/2,2\right)$, fitting
the constants $a,b,\nu$ in order the three first moments of $\sum_{0}^{m}\lambda_{i}Z_{i}^{2}$
and $a+b\chi_{\nu}^{2}$ be equal. This leads us to solve:\begin{equation}
a+b\ \nu=\sum_{0}^{m}\lambda_{i},\quad b^{2}\nu=\sum_{0}^{m}\lambda_{i}^{2},\quad b^{3}\nu=\sum_{0}^{m}\lambda_{i}^{3}\label{app}\end{equation}
Table 1 shows the eigenvalues of the asymptotic covariance matrix
of $T_{m}$ and the corresponding constants, $a$, $b$, and $\nu$
for $m=1,2,3$ and $4$.

Table 2 shows the critical points, obtained by simulation, for the
$T_{m}$ statistics ($m=0,1,2,3$ and $4$) for samples of size $50$,
$100$, $200$, $500$, $1000$ and $2000$, corresponding to the
$90$, $95$ and $99$ percentiles, as well as the values obtained
by simulation of the asymptotic distribution $\left(\ref{p3}\right)$
and the approximation given from $\left(\ref{app}\right)$. The simulations
are all run with $50,000$ samples. It can be seen that the asymptotic
and approximate methods are useful for samples larger than $500$.
These two methods are particularly useful for finding rough p-values.
Note that for the approximate method,\[
\Pr\left\{ T_{m}>t\right\} =\Pr\left\{ \chi_{\nu}^{2}>(t-a)/b\right\} \]
where $a,b,\nu$ are the solutions of $\left(\ref{app}\right)$.

\begin{table}[h]
$$
\begin{array}{|c|c|c|c|c|c|c|c|c|}
\hhline{~--------} 
\omit & \multicolumn{5}{|c|}{\text{EigenValues}} & \multicolumn{3}{c|}{\text{Parameters}}\\
\hhline{~--------} 
\omit&\multicolumn{1}{|c}{\lambda 1}  & \lambda 2  & \lambda 3  & \lambda 4 & \lambda 5  & a  & b  & \nu \\
\hhline{---------} 
T1  & 1.7071  & 0.2929  & -  & -  & -  & 0.2000  & 1.6667  & 1.0800 \\
T2  & 2.2808  & 0.5000  & 0.2192  & -  & -  & 0.4792  & 2.1818  & 1.1554 \\
T3  & 2.7503  & 0.7420  & 0.3104  & 0.1974  & -  & 0.7971  & 2.5758  & 1.2435 \tabularnewline
T4  & 3.1381  & 1.0000  & 0.4241  & 0.2500  & 0.1879  & 1.1323  & 2.8764  & 1.3446 \tabularnewline
\hhline{---------} 
\end{array}
$$
\caption{The eigenvalues of the asymptotic covariance matrix of V and the corresponding
constants for the approximate distribution.}
\end{table}

\subsection{An example}

This analysis is based on the $n=2000$ largest values for the positive
and negative parts of dollar/yen returns, respectively, introduced
in Section 3. The corresponding CV-plots are (a) and (b) in Figure
1. Looking at the CV plot it can be think that exponentiality is accepted
for high order statistics, even in the negative part. In fact when
the sample is small enough always the null hypothesis is accepted.
But looking at the CV plot hundreds of test are done.

Here, the statistic $T_{m}$, for $m=7$, is used; see $\left(\ref{cd}\right)$.
The coefficients of variation over tresholds, $cv_{k}$ , for $k=0,...,7$,
and samples size $n_{k}=n2^{-k}$ are the following: for the positive
part\[
\{0.978,0.959,1.008,1.002,1.018,0.919,1.015,0.968\}\]
and for the negative part\begin{equation}
\{1.088,1.135,1.141,1.111,1.088,1.138,1.16,1.585\}\label{cvs}\end{equation}
The $T_{m}$ statistics and their corresponding p-values are given
by $T_{m}=3.15$ and $p=0.784$, for the positive part; $T_{m}=54.92$
and $p=0.002$, for the negative part. Hence, we accept exponential
tails for the positive part and reject this hypothesis for the negative
part. Note that in the first case we accept exponentiality for a really
large sample, not only the high upper tail of the distribution, and
that our test uses simultaneously eight thresholds. The CV-plot in
Figure 1(b) suggests a constant coefficient of variation greater than
$1$; thus a Pareto distribution can be assumed (Sullo and Rutherford,
1977).

In our analysis we conclude that the tails for the positive part of
the returns decrease exponentially fast. However, for the negative
part we conclude that the tails decrease at a polynomial rate. These
conclusions can be surprising, since by considering the yen denominated
in dollars the positive and negative part change from one to the other.
Note that in these 25 years the price of one dollar went down from
200 yen to 100 yen, more or less. Perhaps this fact and the different
sizes of the two economies can explain the difference between positive
and negative parts. Probably the traders use different strategies
when these two currencies go up or go down. We do not know what the
dollar will do in future years. We believe that if it goes down a
polynomial rates would be correct to measure risks.

\subsection{Comparisons with other inference approaches}

The CV-plot (b) in Figure 1 suggest to model the negative part of
dollar/yen returns by a Pareto distribution. The \emph{generalized
Pareto family of distributions (GPD)} has probability distribution
function, for $\beta>0$,\begin{equation}
F\left(x\right)=1-\left(1+\xi x/\beta\right)^{-1/\xi},\label{pareto}\end{equation}
defined on $x>0$ for $\xi>0$ and defined on $0<x<\beta/\left\vert \xi\right\vert $
for $\xi<0$. The limit case $\xi=0$ corresponds to the exponential
distribution. When $\xi>0$, the GPD is simply the \emph{Pareto distribution}.
In this case the tail function decrease like a power law and the inverse
of the shape parameter, $\xi^{-1}$, is called the \emph{power of
the tail}.

Hence we can estimate the parameters of $\left(\ref{pareto}\right)$\ by
maximum likelihood (ML), using the sample of size $n=2000$ in the
last Example. We find $\hat{\xi}^{-1}=13.473$ and $\hat{\beta}=$
$0.024\ $\ and, the corresponding coefficient of variation is $\hat{c}_{\xi}=1.084$.
Note that this result is not far from $c_{0}=1.088$ in $\left(\ref{cvs}\right)$.
In the same way, estimating the Pareto parameters by ML, from samples
of size $n_{k}$, we find coefficients of variation near $c_{k}$
in $\left(\ref{cvs}\right)$.

The clasical approach from extreme values theory uses \bigskip{}
 the \emph{generalized extreme value distribution}. This distribution
is defined by the cumulative distribution function\begin{equation}
G\left(x\right)=\exp\left[-\left(1+\xi\left(\frac{x-\mu}{\sigma}\right)\right)_{+}^{-1/\xi}\right].\label{gev}\end{equation}

For $\xi>0$ the model\ $\left(\ref{gev}\right)$\ is the Frechet
distribution, for $\xi=0$ the Gumbel distribution and for $\xi<0$
the Weibull distribution, see Embrechts et al. (1997).

Using $\left(\ref{gev}\right)$, with the anual maximums gives the
ML estimation \[
\left(\hat{\mu},\hat{\sigma},\hat{\xi}^{-1}\right)=\left(0.023,0.005,5.485\right)\]
and leads to $\hat{c}_{\xi}=1.255$. The standard error for $\hat{\xi}^{-1}$
has been computed with the inverse of the observed information matrix,
and gives $sd($ $\hat{\xi}^{-1})=5.326$. Hence, the $95\%$ confidence
interval for $\xi^{-1}$ includes the estimation above. However, the
range for $\xi^{-1}$\ is really wide, including distributions with
no finite mean and distributions with compact support.

We conclude that the estimation done with Pareto distribution seems
correct and it agrees with the hypotesis of a coefficient of variation
over thresholds constant. However, the tail estimated with generalized
extreme value distribution looks away of the coefficients of variation
over threshold in $\left(\ref{cvs}\right)$.

\section{Power estimates}

\label{powerestimates}

The $T_{m}$ statistics test simultaneously at several points whether
$CV=1$, though at each new point only one half of the sample of the
previous point is used. Hence, $T_{m}$ statistics are especially
useful for testing exponentiality in the tails, when the exact point
where the tail begins is unknown, avoiding the problem of multiple
comparisons. However, in this Section $T_{m}$ is considered as a
simple test of exponentiality.

Two experiments are conducted. The first one considers as the alternative
distribution the absolute value of the Student distribution (with
degrees of freedom $\nu=1$ to $10)$. In the second case the alternative
distribution is a Pareto distribution. In both cases the empirical
powers of the $T_{m}$ statistics ($m=0,1,2,3$ and $4$) have been
compared with the empirical powers of the empirical coefficient of
variation (D'Agostino and Stephens, 1986) and the tests suggested
by Montfort and Witter (1985) and Smith (1975) as tests against heavy-tailed
alternatives. Every empirical power is estimated running $10,000$
samples and using the critical points of Tables 2 and 3. All the statistics
considered are invariant to changes in scale parameters. Hence, the
powers estimated do not depend on scale parameters under the null
hypothesis of exponentiality or under the alternative distributions.

Montfort and Witter (1985) propose the maximum/median statistic for
testing exponentiality against the GPD. Given a sample $\left\{ X_{i}\right\} $,
let us denote \begin{equation}
MW=Max\left(X_{i}\right)/X_{m}\label{HW}\end{equation}
where $X_{m}$ is the median of the sample.

Smith (1975) and Gel, Miao and Gastwirth (2007) show that powerful
tests of normality against heavy-tailed alternatives are obtained
using the average absolute deviation from the median. The same statistic
suggested by Smith (1975) is used here for testing exponentiality
against heavy-tailed alternatives. Let us denote\begin{equation}
SU=\left[\sum_{i}\left(X_{i}-\bar{X}\right)^{2}/n\right]^{1/2}/\left[\sum_{i}\left\vert X_{i}-X_{m}\right\vert /n\right]\label{SU}\end{equation}
where $\bar{X}$ is the sample mean.

The empirical coefficient of variation statistics is (D'Agostino and
Stephens, 1986)\[
cv=\left[\sum_{i}\left(X_{i}-\bar{X}\right)^{2}/n\right]^{1/2}/\bar{X}.\]

Table 3 shows the critical points for the empirical coefficient of
variation and the statistics $MW$ and $SU$, for samples of size
equal to $50$, $100$, $200$, $500$, $1000$ and $2000$, corresponding
to several quantiles. The simulations are all run with $50,000$ samples.
Note that here two-sided test are considered. This one is the unique
difference between $cv$ and $T_{0}$.

The cumulative distribution function of the Pareto distribution is
\begin{equation}
F\left(x\right)=1-\left(1+\xi x/\psi\right)^{-1/\xi},\label{GPD}\end{equation}
where $\psi>0$ and $\xi>0$ are scale and shape parameters and $x>0$.
The limit case $\xi=0$ corresponds to the exponential distribution.
The parameter $\alpha=1/\xi$ is called the power of the tail.

The probability density function of the Student distribution with
$\nu$ degrees of freedom is\[
t_{\nu}\left(x\right)=\frac{\Gamma\left(\left(\nu+1\right)/2\right)}{\sqrt{\nu\pi}\ \Gamma\left(\nu/2\right)}\left(1+\frac{x^{2}}{\nu}\right)^{-\left(\nu+1\right)/2}\]
Hence, a Student distribution is a distribution of regular variation
with index $\alpha=\nu$. That is, the tails of the Student distribution
are like the Pareto distribution for $\xi=1/\nu$. When $\nu$ tends
to infinity the Student distribution tends to the standard normal
distribution, hence it is a usual alternative when the tails are heavier
than in the normal case. For $\nu=1$ the distribution is also called
the Cauchy distribution. In order to test exponentiality only the
positive part, or equivalently the absolute value, of the Student
distribution is considered. Note that in finance often models with
only three finite moments (infinite kurtosis) are considered; that
corresponds to a Student distribution with $\nu=3$ or $\nu=4$.

Table 4 reports the results for the eight statistics with sample sizes,
$n$, of $50,100,200,500,1000$ and $2000$, at significance level
$5\%$, testing exponentiality against the absolute value of the Student
distribution with degrees of freedom from $\nu=1$ to $10$. Several
overall observations can be made on the basis of these sampling experiments.
First of all, the powers are high for $\nu=1$ (Cauchy distribution)
or $\nu=2$ (unbounded variance) and clearly increase with sample
size for $\nu\geq7$. In most cases $cv$ (or $T_{0}$) is superior
to the other tests. However, its power is poor against some particular
cases. Even for samples of size $2000$ the power is only $38\%$
against the absolute values of the Student distribution $t_{4}$.
This is easily explained since the alternative has coefficient of
variation $CV=1$, as in the null hypothesis of exponentiality. In
this case the powers of $T_{1}$, $T_{2}$ and $T_{3}$ are $96\%$,
$98\%$ and $97\%.$ In general the power of $cv$ is something higher
than $T_{1}$ or $T_{2}$ but in some cases very much lower.

Table 5 reports the results of the eight statistics with sample sizes,
$n$, of $50,100,200,500$ and $1000$, at significance level $5\%$,
testing exponentiality against a Pareto distribution with scale parameter
$\psi=1$ and shape parameters $\xi$ from $0.05$ to $0.5$ with
increments of $0.05$. The Pareto distribution has constant coefficient
of variation, hence the $T_{m}$ statistics do not have any advantage
testing for $CV=1$ at different points. Moreover, at each new point
only one half of the sample of the previous point is used. The overall
observation that can be made on the basis of these sampling experiments
is that again $cv$ (or $T_{0}$) is superior to other tests; this
agrees with the results Castillo and Daoudi (2009). Moreover, other
$T_{m}$ statistics are not far away from $cv$.

The main conclusion is that, though $cv$ is in general a good test,
the $T_{m}$ statistics have a very similar power and clearly improve
the poor power of $cv$ in testing against distributions with coefficient
of variation near $1$, which often appear in finance.

\section{Bibliography}
\begin{enumerate}
\item Ascher, S. (1990). A survey of tests for exponentiality. \emph{Communications
in Statistics: Theory and Methods}, 19, 1811--1825.
\item Balkema, A. and DeHaan, L. (1974). Residual life time at great age.
\emph{Annals of Probability}. 2, 792-804.
\item Coles, S. (2001). \emph{An Introduction to Statistical Modelling of
Extreme Values}. Springer, London.
\item Castillo, J. and Daoudi, J. (2009). Estimation of the generalized
Pareto distribution. \emph{Statistics and Probability Letters}. 79,
684-688.
\item Castillo, J. and Puig, P. (1999). The Best Test of Exponentiality
Against Singly Truncated Normal Alternatives. \emph{Journal of the
American Statistical Association}. 94, 529-532.
\item D'Agostino, R. and Stephens, M.A. (1986). \emph{Goodness-of-Fit Techniques}.
Marcel Dekker, New York.
\item Embrechts, P. Klüppelberg, C. and Mikosch, T. (1997). \emph{Modeling
Extremal Events for Insurance and Finance}. Springer, Berlin.
\item Finkenstadt, B.and Rootzén, H. (edit) (2003). \emph{Extreme values
in Finance, Telecommunications, and the Environment}. Chapman \& Hall.
\item Lee, S., Locke, C. and Spurrier, J. (1980). On a class of tests of
exponentiality. \emph{Technometrics}, 22, 547--554.
\item McNeil, A. Frey, R. and Embrechts, P. (2005). \emph{Quantitative Risk
Management: Concepts, Techniques and Tools}. Princeton UP. New Jersey.
\item Mimoto, N. and Zitikis, R. (2008). The Atkinson index, the Moran statistic
and testing exponentiality. \emph{J. Japan Statist. Soc}. 38, 187--205.
\item Montfort, M. and Witter, J. (1985). Testing exponentiality against
generalizad Pareto distribution. \emph{Journal of Hydrology}, 78,
305-315.
\item Gel, Y., Miao, M. and Gastwirth, J. (2007). Robust directed tests
of normality against heavy-tailed alternatives. \emph{Computational
Statistics \& Data Analysis} 51, 2734-2746.
\item Ghosh, S. and Resnick, S. (2010). A discussion on mean excess plots.
\emph{\ Stochastic Processes and their Applications} 120, 1492-1517.
\item Gupta, R. (1987). On the monotonic properties of the residual variance
and their applications in reliability. \emph{Journal of Statistical
Planning and Inference} 16, 329-335.
\item Gupta, R. and Kirmani, S. (2000). Residual coefficient of variation
and some characterization results. \emph{Journal of Statistical Planning
and Inference} 91, 23-31.
\item Pickands, J. (1975). Statistical inference using extreme order statistics.
\emph{The Annals of Statistics} 3, 119-131.
\item Smith, V. (1975). A Simulation analysis of the Power of Several Test
for Detecting Heavy-Tailed Distributions. \emph{Journal of the American
Statistical Association}, 70, 662-665.
\item Sullo, P. and Rutherford, D. (1977). Characterizations of the Power
Distribution by Conditional Exceedance, in American Statistical Association,
Proceedings of the Business and Economic Statistics Section, Washington,
D.C. 
\end{enumerate}

\begin{table}[h]
\footnotesize{
$$
\begin{array}{|@{\,}c@{\,}|@{\,}c@{\,}|@{\,}c@{\,}|@{\,}c@{\,}|@{\,}c@{\,}|@{\,}c@{\,}|@{\,}c@{\,}|@{\,}c@{\,}|@{\,}c@{\,}|@{\,}c@{\,}|@{\,}c@{\,}|@{\,}c@{\,}|@{\,}c@{\,}|@{\,}c@{\,}|@{\,}c@{\,}|@{\,}c@{\,}|}
\hhline{~---------------} 
\multicolumn{1}{c|@{\,}}{}&\multicolumn{3}{c|@{\,}}{T0} & \multicolumn{3}{c|@{\,}}{T1} & \multicolumn{3}{c|@{\,}}{T2} & \multicolumn{3}{c|@{\,}}{T3} & \multicolumn{3}{c|}{T4}\\
\hline 
\text{\tiny {Sample $(n)$}} &90 &95 &99 &90 &95 &99 &90 &95 &99 &90 &95 &99 &90 &95 &99 \\
\hline 
50 &\hfil 2.19 &3.02 &5.79 &3.59 &4.88 &9.63 &4.73 &6.21 &11.85 &5.51 &7.02 &12.60 &6.11 &7.62 &12.77 \\
100  &2.38  &3.36 &6.43 &4.03 &5.61 &11.30 &5.29 &7.14 &14.45 &6.34 &8.35 &16.27 &7.06 &9.08  
&16.90 \\
200 &2.47 &3.54 &6.46 &4.36 &6.16 &11.48 &5.85 &8.07 &15.58 &7.04 &9.48 &18.80 &8.03 &10.57 &20.32 \\
500 &2.59 &3.71 &6.68 &4.65 &6.52 &11.85 &6.43 &8.85 &16.36 &7.99 &10.72 &19.89 &9.19 &12.24 &22.86 \\
1000 &2.64 &3.74 &6.70 &4.80 &6.65 &11.76 &6.64 &9.21 &16.23 &8.30 &11.37 &20.17 &9.79 &13.16 &23.41 \\
2000 &2.70 &3.83 &6.54 &4.89 &6.84 &11.63 &6.89 &9.39 &15.94 &8.65 &11.63 &19.93 &10.17 &13.55 &23.33 \\
\text{\tiny{Asimptotic}} &2.71 &3.84 &6.63 &4.99 &6.97 &11.62 &7.04 &9.60 &15.98 &8.96 &12.04 &19.69 &10.80 &14.39 &22.88 \\
\text{\tiny{Approximate}} &2.71 &3.84 &6.63 &4.99 &6.93 &11.65 &7.09 &9.67 &15.94 &9.06 &12.18 &19.69 &10.93 &14.49  &23.01 \\\hline
\end{array}
$$}
\caption{The critical points for the $T_{m}$ statistics ($m=0,1,2,3$ and $4$)
for several sample sizes, corresponding to the $90$, $95$ and $99$ percentiles,
as well as the values obtained with the asymptotic distribution and
its approximation.}
\end{table}

\newpage

\begin{table}[h]
\footnotesize{
$$
\begin{array}{|@{\,}c@{\,}|@{\,}c@{\,}|@{\,}c@{\,}|@{\,}c@{\,}|@{\,}c@{\,}|@{\,}c@{\,}|@{\,}c@{\,}|@{\,}c@{\,}|@{\,}c@{\,}|@{\,}c@{\,}|@{\,}c@{\,}|}
\hline 
\text{Sample }(n) &\text{Statistic}  & 0.01  & 0.025  & 0.05  & 0.1  & 0.5  & 0.9  & 0.95  & 0.975  & 0.99 \\
\hline 
20  & CV  & 0.593  & 0.635  & 0.674  & 0.722  & 0.914  & 1.174  & 1.266  & 1.354  & 1.472 \\
50  & CV  & 0.733  & 0.764  & 0.791  & 0.823  & 0.959  & 1.138  & 1.201  & 1.256  & 1.334 \\
100  & CV  & 0.800  & 0.824  & 0.846  & 0.872  & 0.977  & 1.108  & 1.152  & 1.194  & 1.248 \\
200  & CV  & 0.854  & 0.873  & 0.890  & 0.910  & 0.988  & 1.081  & 1.112  & 1.140  & 1.176 \\
500  & CV  & 0.903  & 0.916  & 0.928  & 0.942  & 0.995  & 1.054  & 1.073  & 1.090  & 1.111 \\
1000  & CV  & 0.930  & 0.940  & 0.949  & 0.959  & 0.997  & 1.040  & 1.052  & 1.064  & 1.078 \\
2000  & CV  & 0.950  & 0.957  & 0.964  & 0.971  & 0.999  & 1.028  & 1.037  & 1.044  & 1.053 \\
\hline
20  & HW  & 2.163  & 2.388  & 2.631  & 2.978  & 4.855  & 8.573  & 10.204  & 11.851  & 14.134 \\
50  & HW  & 3.288  & 3.582  & 3.880  & 4.272  & 6.199  & 9.573  & 10.870  & 12.207  & 14.120 \\
100  & HW  & 4.194  & 4.543  & 4.859  & 5.271  & 7.181  & 10.324  & 11.531  & 12.719  & 14.345 \\
200  & HW  & 5.248  & 5.573  & 5.909  & 6.307  & 8.182  & 11.127  & 12.281  & 13.413  & 14.884 \\
500  & HW  & 6.601  & 6.951  & 7.278  & 7.674  & 9.506  & 12.335  & 13.457  & 14.520  & 15.924 \\
1000  & HW  & 7.636  & 7.993  & 8.298  & 8.691  & 10.488  & 13.310  & 14.359  & 15.421  & 16.802 \\
2000  & HW  & 8.701  & 9.030  & 9.344  & 9.730  & 11.489  & 14.269  & 15.287  & 16.308  & 17.674 \\
\hline
20  & SU  & 1.127  & 1.150  & 1.172  & 1.202  & 1.359  & 1.629  & 1.735  & 1.838  & 1.974 \\
50  & SU  & 1.201  & 1.224  & 1.245  & 1.272  & 1.401  & 1.595  & 1.667  & 1.738  & 1.829 \\
100  & SU  & 1.250  & 1.271  & 1.291  & 1.315  & 1.417  & 1.561  & 1.613  & 1.665  & 1.731 \\
200  & SU  & 1.297  & 1.314  & 1.330  & 1.349  & 1.430  & 1.533  & 1.570  & 1.603  & 1.645 \\
500  & SU  & 1.342  & 1.356  & 1.367  & 1.381  & 1.436  & 1.503  & 1.525  & 1.546  & 1.572 \\
1000  & SU  & 1.369  & 1.379  & 1.388  & 1.399  & 1.440  & 1.487  & 1.502  & 1.515  & 1.531 \\
2000  & SU  & 1.390  & 1.397  & 1.404  & 1.411  & 1.441  & 1.474  & 1.484  & 1.493  & 1.503 \\
\hline
\end{array}
$$}
\caption{The critical points for the sampling coefficient of variation ($CV$)
and the statistics MW and SU, for several sample sizes and several
percentiles.}
\end{table}

\medskip

\footnotesize{
\begin{longtable}[c]{|c|c|c|c|c|c|c|c|c|c|} 
\hline 
$\nu$ &$n$ &$cv$ &$HW$ &$SU$ &$T0$ &$T1$ &$T2$ &$T3$ &$T4$\\
\hline 
1 &50 &0.948 &0.933 &0.914 &0.951 &0.940 &0.933 &0.931 &0.930 \\
2 &50 &0.441 &0.400 &0.455 &0.447 &0.465 &0.448 &0.442 &0.436 \\
3 &50 &0.207 &0.163 &0.206 &0.200 &0.222 &0.218 &0.213 &0.207  \\
4 &50 &0.177 &0.120 &0.126 &0.157 &0.147 &0.147 &0.144 &0.140  \\
5 &50 &0.196 &0.126 &0.092 &0.186 &0.139 &0.134 &0.130 &0.127  \\
6 &50 &0.241 &0.151 &0.088 &0.212 &0.154 &0.139 &0.138 &0.135  \\
7 &50 &0.278 &0.180 &0.095 &0.254 &0.173 &0.158 &0.154 &0.150  \\
8 &50 &0.309 &0.202 &0.100 &0.280 &0.193 &0.168 &0.161 &0.158  \\
9 &50 &0.338 &0.221 &0.110 &0.304 &0.208 &0.182 &0.176 &0.172  \\
10 &50 &0.373 &0.247 &0.119 &0.324 &0.226 &0.196 &0.188 &0.183  \\
\hline
1 &100 &0.998 &0.995 &0.993 &0.998 &0.997  &0.997  &0.996 &0.996 \\
2 &100 &0.649 &0.599 &0.684 &0.670 &0.696  &0.685 &0.672 &0.668 \\
3 &100 &0.256 &0.229 &0.307 &0.268 &0.322  &0.336 &0.324 &0.319 \\
4 &100 &0.203 &0.137 &0.151 &0.193 &0.190  &0.200 &0.193 &0.190 \\
5 &100 &0.280 &0.159 &0.112 &0.245 &0.184  &0.178 &0.170 &0.165 \\
6 &100 &0.350 &0.195 &0.112 &0.323 &0.223  &0.200 &0.184 &0.178 \\
7 &100 &0.439 &0.258 &0.136 &0.405 &0.278  &0.236 &0.218 &0.212 \\
8 &100 &0.511 &0.301 &0.158 &0.462 &0.323  &0.268 &0.248 &0.241 \\
9 &100 &0.556 &0.339 &0.182 &0.516 &0.361  &0.298 &0.275 &0.267 \\
10 &100 &0.604 &0.381 &0.206 &0.559 &0.396 &0.326 &0.294 &0.284 \\
\hline
1 &200 &1.000 &1.000 &1.000 &1.000 &1.000 &1.000 &1.000 &1.000  \\
2 &200 &0.883 &0.813 &0.914 &0.891 &0.906 &0.900 &0.891 &0.886  \\
3 &200 &0.360 &0.341 &0.481 &0.376 &0.478 &0.495 &0.492 &0.483  \\
4 &200 &0.248 &0.175 &0.212 &0.237 &0.260 &0.281 &0.286 &0.278  \\
5 &200 &0.393 &0.185 &0.146 &0.346 &0.264 &0.250 &0.247 &0.236  \\
6 &200 &0.557 &0.260 &0.159 &0.512 &0.359 &0.309 &0.290 &0.276  \\
7 &200 &0.682 & 0.345 & 0.217  & 0.646 &0.470 &0.392 &0.349 &0.325  \\
8 &200 &0.766 &0.424 &0.282 &0.743 &0.568 &0.477 &0.427 &0.400  \\
9 &200 &0.825 &0.492 &0.330 &0.798 &0.635 &0.541 &0.487 &0.458  \\
10 &200 &0.870 &0.535 &0.382 &0.846 &0.694 &0.601 &0.543 &0.512 \\
\hline
1 &500 &1.000 &1.000 &1.000 &1.000 &1.000 &1.000 &1.000 &1.000\\
2 &500 &0.996 &0.972 &0.998 &0.994 &0.996 &0.996 &0.995 &0.994\\
3 &500 &0.565 &0.549 &0.776 &0.574 &0.763 &0.781 &0.773 &0.763\\
4 &500 &0.305 &0.231 &0.315 &0.301 &0.429 & 0.482 &0.477 &0.462\\
5 &500 &0.577 &0.199 &0.166  &0.569 &0.507 &0.491 &0.456 &0.427\\
6 &500 &0.818 &0.312 &0.248 &0.804 &0.708 &0.648 &0.596 &0.551\\
7 &500 &0.928  &0.441 &0.396 &0.927 &0.850 & 0.797 &0.745 &0.696 \\
8 &500 &0.972 &0.548 &0.534 &0.966 &0.922 & 0.887 &0.847 &0.814  \\
9 &500 &0.988 &0.639 &0.648 & 0.987 &0.962  & 0.934 &0.906 &0.877 \\
10 &500 &0.994 &0.709 &0.741 &0.994 &0.980 &  0.964 &0.944 &0.924 \\
\hline
1 &1000 &1.000 &1.000 &1.000 &1.000 &1.000 &1.000 &1.000 &1.000  \\
2 &1000 &1.000 &0.998 &1.000 &1.000 &1.000 &1.000 &1.000 &1.000 \\
3  &1000 &0.758 &0.730 &0.952 &0.771 &0.945 &0.953 &0.950 &0.942  \\
4 &1000 &0.336 &0.294 &0.463 &0.346 &0.703 &0.759 &0.739 &0.714  \\
5 &1000 &0.745 &0.217 &0.195 &0.738 &0.824 &0.832 &0.788 &0.742  \\
6 &1000 &0.951 &0.331 &0.333 &0.950 &0.956 &0.951 &0.922 &0.890  \\
7 &1000 &0.991 &0.475 &0.579 &0.991 &0.992 &0.990 &0.982 &0.969  \\
8 &1000 &0.998 &0.619 &0.770 &0.998 &0.999 &0.998 &0.996 &0.992 \\
9 &1000 &1.000 &0.715 &0.880 &0.999 &1.000 &0.999 &0.999 &0.998 \\
10 &1000 &1.000 &0.784 &0.935 &1.000 &1.000 &1.000 &1.000 &1.000   \\
\hline
1 &2000 &1.000 &1.000 &1.000 &1.000 &1.000 & 1.000 &1.000 &1.000\\
2 &2000 &1.000 &1.000 &1.000 &1.000 &1.000 & 1.000 &1.000 &1.000\\
3 &2000 &0.937 &0.891 &0.998 &0.939 &0.998 & 0.999 &0.999 &0.998 \\
4 &2000 &0.383 &0.414 &0.679 &0.373 &0.962 & 0.976 &0.966 &0.951 \\
5 &2000 &0.882 &0.232 &0.230 &0.888 &0.995 & 0.995 &0.993 &0.986  \\
6 &2000 &0.991 &0.321 &0.484 &0.992 &1.000  & 1.000 &1.000 &0.999  \\
7 &2000 &0.999 &0.506 &0.811 &0.999 &1.000 & 1.000 &1.000 &1.000  \\
8 &2000 &1.000 &0.658 &0.951 &1.000 &1.000 & 1.000 &1.000 &1.000  \\
9 &2000 &1.000 &0.760 &0.987 &1.000 &1.000 & 1.000 &1.000 &1.000 \\
10 &2000  &1.000 &0.834 &0.997 &1.000 &1.000 &1.000 &1.000 &1.000\\
\hline
\end{longtable}}

\noindent
\normalsize{Figure 4: Power of eight statistics with several sample sizes, $n$, at significance
level of 5{\%}, testing exponentiality against a Student distribution
with degrees of freedom from~$1$ to $10$. The power is estimated using
$10,000$ samples.}

\medskip

\footnotesize{
\begin{longtable}[c]{|c|c|c|c|c|c|c|c|c|c|}
\hline 
$\theta$  & $n$  & $cv$  & $HW$  & $SU$  & $T0$  & $T1$  & $T2$  & $T3$  & $T4$ \\
\hline 
0.05  & 50  & 0.078  & 0.073  & 0.072  & 0.079  & 0.080  & 0.075  & 0.074  & 0.071 \\
0.10  & 50  & 0.136  & 0.119  & 0.112  & 0.137  & 0.136  & 0.124  & 0.117  & 0.114 \\
0.15  & 50  & 0.212  & 0.189  & 0.175  & 0.223  & 0.215  & 0.200  & 0.193  & 0.189 \\
0.20  & 50  & 0.302  & 0.273  & 0.249  & 0.317  & 0.292  & 0.267  & 0.257  & 0.252 \\
0.25  & 50  & 0.396  & 0.356  & 0.313  & 0.416  & 0.387  & 0.359  & 0.348  & 0.342 \\
0.30  & 50  & 0.493  & 0.452  & 0.388  & 0.494  & 0.458  & 0.429  & 0.419  & 0.414 \\
0.35  & 50  & 0.577  & 0.528  & 0.453  & 0.594  & 0.552  & 0.518  & 0.505  & 0.499 \\
0.40  & 50  & 0.654  & 0.609  & 0.534  & 0.661  & 0.619  & 0.589  & 0.578  & 0.574 \\
0.45  & 50  & 0.729  & 0.685  & 0.604  & 0.739  & 0.696  & 0.667  & 0.659  & 0.655 \\
0.50  & 50  & 0.784  & 0.742  & 0.654  & 0.799  & 0.753  & 0.727  & 0.720  & 0.717 \\
\hline
0.05  & 100  & 0.094  & 0.088  & 0.086  & 0.100  & 0.099  & 0.097  & 0.095  & 0.092 \\
0.10  & 100  & 0.185  & 0.162  & 0.159  & 0.210  & 0.197  & 0.186  & 0.178  & 0.177 \\\
0.15  & 100  & 0.330  & 0.271  & 0.269  & 0.356  & 0.333  & 0.311  & 0.293  & 0.289 \\
0.20  & 100  & 0.476  & 0.392  & 0.376  & 0.505  & 0.467  & 0.443  & 0.420  & 0.413 \\
0.25  & 100  & 0.622  & 0.537  & 0.502  & 0.648  & 0.603  & 0.568  & 0.550  & 0.542 \\
0.30  & 100  & 0.744  & 0.652  & 0.619  & 0.760  & 0.715  & 0.687  & 0.667  & 0.662 \\
0.35  & 100  & 0.831  & 0.746  & 0.707  & 0.841  & 0.801  & 0.777  & 0.760  & 0.757 \\
0.40  & 100  & 0.892  & 0.829  & 0.786  & 0.897  & 0.863  & 0.842  & 0.829  & 0.825 \\
0.45  & 100  & 0.933  & 0.881  & 0.846  & 0.943  & 0.916  & 0.898  & 0.888  & 0.888 \\
0.50  & 100  & 0.958  & 0.920  & 0.885  & 0.964  & 0.945  & 0.933  & 0.926  & 0.923 \\
\hline
0.05  & 200  & 0.131  & 0.101  & 0.114  & 0.135  & 0.131  & 0.124  & 0.120  & 0.116 \\
0.10  & 200  & 0.299  & 0.219  & 0.236  & 0.329  & 0.297  & 0.277  & 0.263  & 0.255 \\
0.15  & 200  & 0.533  & 0.393  & 0.428  & 0.560  & 0.513  & 0.478  & 0.454  & 0.440 \\
0.20  & 200  & 0.743  & 0.571  & 0.613  & 0.759  & 0.709  & 0.674  & 0.651  & 0.636 \\
0.25  & 200  & 0.865  & 0.714  & 0.747  & 0.879  & 0.833  & 0.803  & 0.784  & 0.772 \\
0.30  & 200  & 0.941  & 0.832  & 0.855  & 0.949  & 0.924  & 0.904  & 0.892  & 0.885 \\
0.35  & 200  & 0.977  & 0.910  & 0.925  & 0.979  & 0.963  & 0.953  & 0.944  & 0.941 \\
0.40  & 200  & 0.990  & 0.951  & 0.958  & 0.992  & 0.986  & 0.981  & 0.977  & 0.975 \\
0.45  & 200  & 0.997  & 0.979  & 0.982  & 0.998  & 0.996  & 0.994  & 0.992  & 0.991 \\
0.50  & 200  & 0.999  & 0.990  & 0.992  & 0.999  & 0.998  & 0.997  & 0.997  & 0.996 \\
\hline
0.05  & 500  & 0.215  & 0.133  & 0.175  & 0.235  & 0.217  & 0.201  & 0.190  & 0.180 \\
0.10  & 500  & 0.578  & 0.328  & 0.461  & 0.610  & 0.567  & 0.522  & 0.487  & 0.467 \\
0.15  & 500  & 0.875  & 0.592  & 0.752  & 0.887  & 0.841  & 0.802  & 0.775  & 0.757 \\
0.20  & 500  & 0.976  & 0.794  & 0.919  & 0.977  & 0.958  & 0.942  & 0.930  & 0.920 \\
0.25  & 500  & 0.996  & 0.919  & 0.978  & 0.997  & 0.994  & 0.989  & 0.985  & 0.982 \\
0.30  & 500  & 1.000  & 0.973  & 0.995  & 1.000  & 0.999  & 0.999  & 0.999  & 0.998 \\
0.35  & 500  & 1.000  & 0.994  & 0.999  & 1.000  & 1.000  & 1.000  & 1.000  & 1.000 \\
0.40  & 500  & 1.000  & 0.998  & 1.000  & 1.000  & 1.000  & 1.000  & 1.000  & 1.000 \\
0.45  & 500  & 1.000  & 1.000  & 1.000  & 1.000  & 1.000  & 1.000  & 1.000  & 1.000 \\
0.50  & 500  & 1.000  & 1.000  & 1.000  & 1.000  & 1.000  & 1.000  & 1.000  & 1.000 \\
\hline
0.05  & 1000  & 0.358  & 0.159  & 0.281  & 0.386  & 0.350  & 0.315  & 0.293  & 0.272 \\
0.10  & 1000  & 0.851  & 0.443  & 0.728  & 0.868  & 0.824  & 0.780  & 0.744  & 0.717 \\
0.15  & 1000  & 0.990  & 0.745  & 0.954  & 0.992  & 0.980  & 0.970  & 0.959  & 0.952 \\
0.20  & 1000  & 1.000  & 0.920  & 0.996  & 1.000  & 1.000  & 0.999  & 0.998  & 0.997 \\
0.25  & 1000  & 1.000  & 0.983  & 1.000  & 1.000  & 1.000  & 1.000  & 1.000  & 1.000 \\
0.30  & 1000  & 1.000  & 0.998  & 1.000  & 1.000  & 1.000  & 1.000  & 1.000  & 1.000 \\
0.35  & 1000  & 1.000  & 1.000  & 1.000  & 1.000  & 1.000  & 1.000  & 1.000  & 1.000 \\
0.40  & 1000  & 1.000  & 1.000  & 1.000  & 1.000  & 1.000  & 1.000  & 1.000  & 1.000 \\
0.45  & 1000  & 1.000  & 1.000  & 1.000  & 1.000  & 1.000  & 1.000  & 1.000  & 1.000 \\
0.50  & 1000  & 1.000  & 1.000  & 1.000  & 1.000  & 1.000  & 1.000  & 1.000  & 1.000 \\
\hline
\end{longtable}}

\noindent
\normalsize{Figure 5: Power of the eight statistics with several sample sizes, $n$, at significance
level 5{\%}, testing exponentiality against a Pareto distribution with
scale parameter~$1$ and shape parameters from~$0.05$ to~$0.5$ ($+0.05$). The
power is estimated using $10,000$~samples.}

\end{document}